# Transport Anisotropy in One-dimensional Graphene Superlattice in the High Kronig-Penney Potential Limit

Tianlin Li,[1] Hanying Chen,[1] Kun Wang,[1] Yifei Hao,[1] Le Zhang,[1] Kenji Watanabe,[2] Takashi Taniguchi,[3] Xia Hong[1,*]

[1] *Department of Physics and Astronomy and Nebraska Center of Materials and Nanoscience, University of Nebraska, Lincoln, Nebraska 68588, USA*

[2] *Research Center for Electronic and Optical Materials, National Institute for Materials Science, 1-1 Namiki, Tsukuba 305-0044, Japan*

[3] *Research Center for Materials Nanoarchitectonics, National Institute for Materials Science, 1-1 Namiki, Tsukuba 305-0044, Japan*

[*] Email: xia.hong@unl.edu

**Abstract**

One-dimensional graphene superlattice subjected to strong Kronig-Penney (KP) potential is promising for achieving electron lensing effect, while previous studies utilizing the modulated dielectric gates can only yield a moderate, spatially dispersed potential profile. Here, we realize high KP potential modulation of graphene via nanoscale ferroelectric domain gating. Graphene transistors are fabricated on $PbZr_{0.2}Ti_{0.8}O_3$ back-gates patterned with periodic, 100-200 nm wide stripe domains. Due to band reconstruction, the h-BN top-gating induces satellite Dirac points in samples with current along the superlattice vector $\hat{s}$, a feature absent in samples with current perpendicular to $\hat{s}$. The satellite Dirac point position scales with the superlattice period ($L$) as $\propto L^\beta$, with $\beta$ = -1.18±0.06. These results can be well explained by the high KP potential scenario, with the Fermi velocity perpendicular to $\hat{s}$ quenched to about 1% of that for pristine graphene. Our study presents a promising material platform for realizing electron supercollimation and investigating flat band phenomena.

Two-dimensional (2D) van der Waals materials subjected to artificially designed superlattice (SL) potential modulation are versatile platforms exhibiting a rich variety of emergent phenomena [1-3], including band reconstruction and Brillouin zone folding [4-12], correlation driven Mott transitions [13], superconductivity [14], magnetism [15, 16], ferroelectricity [17, 18], and topological orders [19]. Among them, the one-dimensional (1D) graphene superlattice (GSL) has gained considerable research interests [1] as the strong anisotropy and flattening of energy bands [6, 7, 20] can lead to electron supercollimation effect [21, 22] as well as correlated states. While 1D GSL has been intensively investigated theoretically [1, 6, 20-25], only few experimental demonstrations have been reported, where the SL potential is generated via a dielectric gate with either nanoscale electrode arrays [7] or periodic thickness modulation [11]. The modulated-dielectric-gating approach has two intrinsic limitations. First, the doping capacity of conventional dielectrics such as $Al_2O_3$ and $SiO_2$ is typically $<10^{13}$ cm$^{-2}$, which imposes only moderate KP potential in the graphene channel. The resulting electronic structure, even though distorted from the isotropic configuration, is not viable for hosting the electron lensing effect. Second, due to the finite gate thickness, the locally applied gate bias becomes spatially dispersed when mapped on the conducting channel, which serves as a global electrode [Fig. 1(a)]. To date, 1D GSL modulation in the high KP potential limit has never been achieved experimentally.

A promising material scheme to overcome these challenges is to exploit a ferroelectric gate with periodically patterned domain structures [26], utilizing the nonvolatile, switchable polarization to induce the SL potential modulation in graphene. The ferroelectric field effect has previously been adopted to induce nonvolatile modulation of the resistance and quantum transport in graphene [27, 28]. Combining it with nanoscale domain patterning further enables the design of reconfigurable functionalities [29-32] and directional conduction paths [33] in a 2D channel. For ferroelectrics such as oxide Pb(Zr,Ti)$O_3$ [34] and copolymer P(VDF-TrFE) [33], the remnant polarization corresponds to a high 2D carrier density that well exceeds $10^{13}$ cm$^{-2}$. Since the surface bound charge of ferroelectric domains is in direct contact with the 2D channel, the induced density variation in graphene across a sharp domain wall (DW) can produce a step-like potential change [Fig. 1(a)].

In this work, we report the first realization of 1D GSL in the high KP potential limit via nanoscale domain patterning in a ferroelectric PbZr$_{0.2}$Ti$_{0.8}$O$_3$ (PZT) gate [Fig. 1(b)]. Monolayer graphene field effect transistors (FETs) are fabricated on prepatterned periodic stripe domains of PZT, with the period $L$ varying from 200 to 300 nm. The polarization reversal of PZT shifts the Fermi level ($E_F$) of graphene by up to 0.9 eV. For current along and perpendicular to the SL vector $ŝ$, the GSL samples exhibit strong transport anisotropy. Satellite Dirac points (DPs) emerge in the longitudinal resistance along $ŝ$ and evolve into multiple Landau fan branches in high magnetic fields, a feature absent for resistance perpendicular to $ŝ$. The average carrier density interval between consecutive DPs scales with the SL period as $\propto L^\beta$, with $\beta = -1.18\pm0.06$, which can be well described by the band structures of 1D GSL under high KP potential. The Fermi velocity perpendicular to $ŝ$ is quenched to about 1% of that for pristine graphene, paving the path for designing the electron lensing effect.

We work with 50 nm epitaxial (001) PZT thin films deposited on 10 nm $La_{0.67}Sr_{0.33}MnO_3$ (LSMO) buffered $SrTiO_3$ (STO) substrates, with the polar axis of PZT along surface normal. The PZT films have high crystallinity and surface roughness of 4-5 Å. Piezoresponse force microscopy (PFM) switching hysteresis shows robust ferroelectric switching with coercive voltages of +1.2/-2.5 V for the polarization down ($P_{down}$) and up ($P_{up}$) states [Fig. 1(c)]. The details of film growth and characterization can be found in the Supplemental Material (SM) Section 1 [35]. In ambient condition, the bound charge of PZT is fully screened prior to graphene transfer and does not induce doping in graphene at 300 K [26, 34]. Upon cooling, on the other hand, the ferroelectric polarization increases due to the pyroelectric effect [28]. We characterize the polarization doping in graphene on PZT using a Hall bar device (SM Section 2) [35]. Figure 1(d) shows the temperature dependence of the converted polarization with respect to the 300 K value, $\Delta P = |P-P(300\ K)|$, which yields a polarization increase of about ~2.5 μC cm$^{-2}$ at 2 K. Switching PZT polarization (2$\Delta P$) can thus lead to a 2D carrier density of up to $3\times10^{13}$ cm$^{-2}$ in graphene.

To fabricate the 1D GSL, we prepattern Cr/Au (2 nm/10 nm) electrodes on PZT in the four-point configuration. Periodic stripe domains are written between the voltage probes by conductive atomic force microscopy in two orientations, either parallel or perpendicular to the current path. The width of the $P_{down}$ ($P_{up}$) domains is 100 nm (100-200 nm), yielding a SL period of $L$ = 200-300 nm [Fig. 1(e)] (SM Section 3) [35]. Monolayer graphene flakes with h-BN top-layers (25-40 nm) are transferred on PZT using the dry transfer approach [36], in direct contact with the prepatterned domain structures and Au electrodes [Fig. 1(f)]. Cr/Au (10 nm/50 nm) are then deposited on h-BN as the global top-gate electrode [Fig. 1(g)]. Magnetotransport measurements are performed in a Quantum Design Physical Property Measurement System using standard lock-in technique with 50 nA current. The results are based on six GSL samples with current along the SL vector $\hat{s}$ (denoted as D1-D6) and one GSL sample with current perpendicular to $\hat{s}$ (denoted as D7), as summarized in Supplemental Table 1 [35]. PFM studies show that the domain structure is robust against sample fabrication and electrical measurements (SM Section 4) [35].

Figure 2 shows the sheet resistance $R_\square$ as a function of top-gate-induced electron doping $\delta n$ at 2 K for two GSL samples, D1 with $L = 205\pm6$ nm [Fig. 2(a)] and D7 with $L = 199\pm6$ nm [Fig. 2(b)]. As $\hat{s}$ is along $x$-axis in the laboratory coordinates, we denote $R_\square$ for these two samples as $R_{xx}$ and $R_{yy}$, respectively. Here $\delta n = \alpha V_g$, where $V_g$ is the top-gate voltage and $\alpha = \varepsilon_r \varepsilon_0 / ed$ is the gating efficiency, with $\varepsilon_r = 3.76$ the dielectric constant of h-BN [37], $\varepsilon_0$ the vacuum permittivity, $e$ the elementary charge, and $d$ the thickness of h-BN (SM Section 5) [35]. For comparison, we also show $R_\square(\delta n)$ for a pristine graphene sample [Fig. 2(a)]. The pristine sample exhibits a single peak at the charge neutral point (CNP), with the field effect mobility $\mu_{FE}$ of about 20,000 cm$^2$V$^{-1}$s$^{-1}$ for holes and 12,000 cm$^2$V$^{-1}$s$^{-1}$ for electrons. For the doping level of interest in this study ($|n| = $ 2-8×10$^{12}$ cm$^{-2}$), this corresponds to a mean free path of about 200-660 nm, confirming that the SL induced band reconstruction is viable at the chosen SL period (200-300 nm). For sample D1, $R_{xx}(\delta n)$ shows two more satellite peaks symmetrically displaced from the main Dirac point, locating at $\delta n = -1.34\times10^{12}$ cm$^{-2}$ and $1.73\times10^{12}$ cm$^{-2}$ [Fig. 2(a)]. This carrier density level is more than an order of magnitude smaller than that expected for pyroelectric polarization doping, excluding inhomogeneous doping as the origin of the emergent peaks. For sample D7, $R_{yy}(\delta n)$ exhibits a single peak [Fig. 2(b)], similar to that of the pristine graphene.

The transport anisotropy between $R_{xx}$ and $R_{yy}$ is a direct manifestation of the emergent 1D GSL [6, 7, 11] induced by the periodic ferroelectric polarization gating effect. At 2 K, the pyroelectric effect induced $\Delta P$ generates a KP-type potential with an equivalent amplitude of about $V_0 = 2\hbar v_F \sqrt{\pi \Delta P(2\text{ K})/e} = 0.9$ eV [Fig. 1(a)], where $v_F = 10^8$ cm s$^{-1}$ is the Fermi velocity of graphene. The band structure is subsequentially renormalized, with the Fermi velocity unchanged along $\hat{s}$ ($v_x$) and highly suppressed perpendicular to $\hat{s}$ ($v_y$) [22]. The extra peaks observed in $R_{xx}(\delta n)$ correspond to the emergent DPs at the band crossing of the highly folded Brillouin zone [1, 6]. The positions of the satellite DPs do not vary significantly from 1.8 to 10 K (SM Fig. S7) [35], consistent with $\Delta P(T)$ of PZT [Fig. 1(d)]. In contrast, $R_{yy}(\delta n)$ only exhibits a single resistance peak at the original CNP due to the suppressed Klein tunneling along the stripe domains [6, 23, 38, 39]. The broadening of the peak compared with pristine graphene can be attributed to the suppressed relaxation time associated with SL-enhanced density of states or a mixture of the $R_{xx}$ component due to small misalignment of the current with respect to $y$-direction. Similar broadening has been reported previously in GSL fabricated via modulated dielectric gates with increasing KP potential [11].

We then investigate the transport anisotropy of GSL in magnetic fields [7, 11]. Figure 3(a) shows $R_{xx}$ versus $\delta n$ and magnetic field $B$ for sample D1, where we observe Shubnikov de Hass oscillations associated with three sets of Landau fan structures (SM Section 6) [35]. The central Landau fan branches out from the original DP, while the two satellite fans emanate from $\delta n$ = -1.41×10$^{12}$ cm$^{-2}$ and 2.16×10$^{12}$ cm$^{-2}$, consistent with the emergent DPs at zero field [Fig. 2(a)]. In Fig. 3(b), we model the Landau fan structures using $n = \nu eB/h + n_{\text{DP}}$, with $\nu = 4l + 2$ the filling factors, $l$ the Landau level index, $h$ the Plank constant, and $n_{\text{DP}}$ the carrier density of the associated DP position. As shown in Fig. 3(a), the simulated Landau fans are in excellent agreement with the experimental $R_{xx}$ data. Figure 3(c) shows the $R_{yy}$ data of sample D7, which possesses a similar SL period. As expected, it exhibits only one set of Landau fan emanating from the original DP [Fig. 3(c)-(d)]. Furthermore, at high magnetic field, $R_{xx}$ is more than one order of magnitude higher than $R_{yy}$, similar as that reported in GSLs with modulated dielectric gates, which has been attributed to the highly localized electron wavefunction along $\hat{s}$ in magnetic fields [11].

The emergent satellite DPs are observed in $R_{xx}$ taken on all six GSL samples with current along $\hat{s}$ (Supplemental Fig. S9) [35]. We adopt multiple-peak fits to $R_{xx}(\delta n)$ to identify the order of extra DPs (SM Section 7) [35]. To minimize the error in calculating the average carrier density interval between two consecutive DPs $\Delta n_{\text{DP}}$, we select two most prominent peaks at high doping levels, with one from the hole branch (order $N_h$ at $n_{\text{DP,h}}$) and one from the electron branch (order $N_e$ at $n_{\text{DP,e}}$), and deduce $\Delta n_{\text{DP}} = (n_{\text{DP,e}} - n_{\text{DP,h}})/(N_e - N_h)$. Figure 4 shows $\Delta n_{\text{DP}}$ as a function of $L$, which can be well described by:

$$\Delta n_{\text{DP}} = AL^\beta, \quad \text{with } \beta = -1.18 \pm 0.06. \tag{1}$$

Here $A$ is the proportionality constant. To understand the $L$-dependence of $\Delta n_{\text{DP}}$, we model the band structures of 1D GSL at moderate and high KP potentials. For 1D GSL under the long wavelength approximation, the Hamiltonian can be written as $H = v_F \vec{\sigma} \cdot \hat{p} + V_{1D}(x)$, with $\vec{\sigma}$ the Pauli matrices vector, $\hat{p}$ the momentum operator, and $V_{1D}(x)$ the 1D scalar KP potential [1, 6, 24]. Assuming the widths of the well and barrier regions are $W_w$ and $W_b$, respectively [Fig. 1(a)], we deduce the dispersion relation $E(k_x, k_y)$ using the transfer matrix method:

$$\cos(k_x L) = \cos\left(\lambda_w \frac{W_w}{L}\right) \cos\left(\lambda_b \frac{W_b}{L}\right) - G \sin\left(\lambda_w \frac{W_w}{L}\right) \sin\left(\lambda_b \frac{W_b}{L}\right), \qquad (2)$$

with $\lambda_{w,b} = \left(\epsilon_{w,b}^2 - k_y^2 L^2\right)^{\frac{1}{2}}$, $G = \frac{\epsilon_w \epsilon_b - k_y^2 L^2}{\lambda_w \lambda_b}$, $\epsilon_w = \epsilon + u\frac{W_w}{L}$, and $\epsilon_b = \epsilon - u\frac{W_b}{L}$. Here $\epsilon_{w,b}$ can be viewed as the dimensionless effective energy of the wavefunction in the well or barrier region, which is given by the dimensionless dispersion $\epsilon = \frac{EL}{\hbar v_F}$ modified by the scaled dimensionless KP potential $u = \frac{V_0 L}{\hbar v_F}$.

We first consider the case for the moderate KP potential as those generated by the modulated dielectric gates [7, 11]. Figure 5(a) shows the modeled band structures for $u = 9\pi$, where new bands emerge along $\hat{s}$ with periodic band crossing points locating at $k_x = N\pi/L$ for the $N$th reconstructed band. The isopotential contours show very complicated features, hosting multiple satellite DPs in one band [Fig. 5(b)]. The Fermi velocity becomes anisotropic, with $v_x$ remaining unchanged and $v_y$ suppressed to about $0.1 v_F$ at the original DP. With increasing $u$, $v_y$ is oscillatorily damped following $v_y = v_F \sin(u/4)/(u/4)$ [6]. For our 1D GSL generated by ferroelectric domains, the SL period $L = 200$ nm corresponds to a high $u \sim 90\pi$, which quenches $v_y$ to about $0.01\, v_F$. As shown in Fig. 5(c), the SL energy band possesses a highly flattened energy dispersion $E \approx \pm \hbar v_F (k_x + 2N\pi/L)$.

We then evaluate the relation between the $E_F$ and $n$ for the SL band structure at the high $u$ limit. From Luttinger theorem, the carrier density is given by the surface area of Fermi surface. As shown in Fig. 5(d), the isopotential contours at $u = 90\pi$ closely resemble rectangles that extend to the entire Brillouin zone. The enclosed area of each isopotential contour can thus be approximated as $\alpha k_x$, with $\alpha$ on the order of the size of graphene Brillouin zone. Given that $n = \alpha k_x/\pi^2$, we obtain $E_F = \hbar v_F n \pi^2/\alpha$. For the satellite DPs located at the $N$th band crossing, $E = \hbar v_F \frac{N\pi}{L}$ and the corresponding doping level is $\Delta n_{DP} = N\alpha/\pi L \propto 1/L$. The inversely proportional relation between $\Delta n_{DP}$ and $L$ closely resembles our experimental results (Fig. 4 and Eq. 1). In sharp contrast, in a moderate KP potential, the extra DP position exhibits an opposite trend and increases with the SL period $L$ at a given KP potential $V_0$ [11]. Setting the fitted proportionality constant in Eq. 1 as $A = \alpha/\pi$, we obtain $\alpha = 26.2$ nm$^{-1}$, which is in excellent agreement with the Brillouin zone size of graphene $2\pi/a_0 = 25.5$ nm$^{-1}$, further confirming that our 1D GSL samples are in the high KP potential limit.

The extremely flattened SL band associated with the high KP potential, on the other hand, can also limit the observation of extra DPs, as the enhanced density of states broadens the resistance peaks [11]. Compared with the multi-satellite DPs observed in 1D GSL subjected to moderate KP potential, *e.g.*, those generated by modulated dielectric gate [7, 11], our samples only exhibit a couple of extra peaks in $R_{xx}(V_g)$, which reflects the flattened dispersion at certain band crossings [Fig. 5(c)]. There are additional factors that can compromise the observation of emergent DPs. First, the ferroelectric DWs are intrinsically rough [40], which leads to a randomness in the SL period $L$ and perturbs the SL band reconstruction [41]. It has been shown theoretically that the transport signature of extra DPs can be damped by ~90% at 5% randomness in the high KP potential limit [42]. Second, the DW roughness for PZT on LSMO is about 6 nm (SM Section 1) [35]. To minimize its impact, we work with SL period $L \geq 200$ nm. Even though the typical mean free path of our 1D GSL samples is larger than $L$, these two length scales become comparable at lower carrier density, which may blur the ideal SL band structure and lead to decoherent transport. Third, we work with finite numbers of the stripe domains (35-50) (SM Section 3) [35], and the convoluted effect of finite SL modulation also broadens the reconstructed band. On the other hand, the limitations associated with a PZT back-gate, including the compromised KP potential due to interfacial screening charges and the non-programmable domain structure after sample fabrication, can be overcome by adopting a suspended PZT membrane top-gate [32].

In conclusion, we realize 1D GSL via periodic ferroelectric stripe domains patterned in a PZT back-gate. The samples exhibit high transport anisotropy between the current directions along and perpendicular to the SL vector, with satellite DPs emerging in the former configuration due to Brillouin zone folding. The scaling behavior of the carrier density interval between consecutive DPs with the SL period can be well modeled by the reconstructed band structure subject to high KP potential. The ferroelectric domain controlled 1D GSL presents a promising material platform for studying electron supercollimation as well as designing collective phenomena associated with flat bands, including magnetism, superconductivity, and topological sub-bands.

**Acknowledgement**

The authors thank Wuzhang Fang, Dawei Li, Qiuchen Wu, and Yuhan Zhang for the valuable discussions, and Anandakumar Sarella for technical assistance. This work was supported by the U.S. Department of Energy (DOE), Office of Science, Basic Energy Sciences (BES), under Award No. DE-SC0016153, NSF EPSCoR RII Track-1: Emergent Quantum Materials and Technologies (EQUATE), under Award No. OIA-2044049, and Nebraska Center for Energy Sciences Research (NCESR). K.W. and T.T. acknowledge support from the JSPS KAKENHI (Grant Numbers 20H00354, 21H05233 and 23H02052) and World Premier International Research Center Initiative (WPI), MEXT, Japan. The research was performed, in part, in the Nebraska Nanoscale Facility: National Nanotechnology Coordinated Infrastructure, the Nebraska Center for Materials and Nanoscience, which are supported by NSF ECCS: 2025298, and the Nebraska Research Initiative.

[7]   S. Dubey, V. Singh, A. K. Bhat, P. Parikh, S. Grover, R. Sensarma, V. Tripathi, K. Sengupta, and M. M. Deshmukh, Tunable Superlattice in Graphene To Control the Number of Dirac Points, Nano Letters **13**, 3990 (2013).

[8]   C. R. Dean, L. Wang, P. Maher, C. Forsythe, F. Ghahari, Y. Gao, J. Katoch, M. Ishigami, P. Moon, M. Koshino, T. Taniguchi, K. Watanabe, K. L. Shepard, J. Hone, and P. Kim, Hofstadter's butterfly and the fractal quantum Hall effect in moiré superlattices, Nature **497**, 598 (2013).

[9]   B. Hunt, J. D. Sanchez-Yamagishi, A. F. Young, M. Yankowitz, B. J. LeRoy, K. Watanabe, T. Taniguchi, P. Moon, M. Koshino, P. Jarillo-Herrero, and R. C. Ashoori, Massive Dirac Fermions and Hofstadter Butterfly in a van der Waals Heterostructure, Science **340**, 1427 (2013).

[10]  L. A. Ponomarenko, R. V. Gorbachev, G. L. Yu, D. C. Elias, R. Jalil, A. A. Patel, A. Mishchenko, A. S. Mayorov, C. R. Woods, J. R. Wallbank, M. Mucha-Kruczynski, B. A. Piot, M. Potemski, I. V. Grigorieva, K. S. Novoselov, F. Guinea, V. I. Fal'ko, and A. K. Geim, Cloning of Dirac fermions in graphene superlattices, Nature **497**, 594 (2013).

[11]  Y. Li, S. Dietrich, C. Forsythe, T. Taniguchi, K. Watanabe, P. Moon, and C. R. Dean, Anisotropic band flattening in graphene with one-dimensional superlattices, Nature Nanotechnology **16**, 525 (2021).

[12]  C. Forsythe, X. Zhou, K. Watanabe, T. Taniguchi, A. Pasupathy, P. Moon, M. Koshino, P. Kim, and C. R. Dean, Band structure engineering of 2D materials using patterned dielectric superlattices, Nature Nanotechnology **13**, 566 (2018).

[13]  Y. Cao, V. Fatemi, A. Demir, S. Fang, S. L. Tomarken, J. Y. Luo, J. D. Sanchez-Yamagishi, K. Watanabe, T. Taniguchi, E. Kaxiras, R. C. Ashoori, and P. Jarillo-Herrero, Correlated insulator behaviour at half-filling in magic-angle graphene superlattices, Nature **556**, 80 (2018).

[14]  Y. Cao, V. Fatemi, S. Fang, K. Watanabe, T. Taniguchi, E. Kaxiras, and P. Jarillo-Herrero, Unconventional superconductivity in magic-angle graphene superlattices, Nature **556**, 43 (2018).

[15]  A. L. Sharpe, E. J. Fox, A. W. Barnard, J. Finney, K. Watanabe, T. Taniguchi, M. A. Kastner, and D. Goldhaber-Gordon, Emergent ferromagnetism near three-quarters filling in twisted bilayer graphene, Science **365**, 605 (2019).

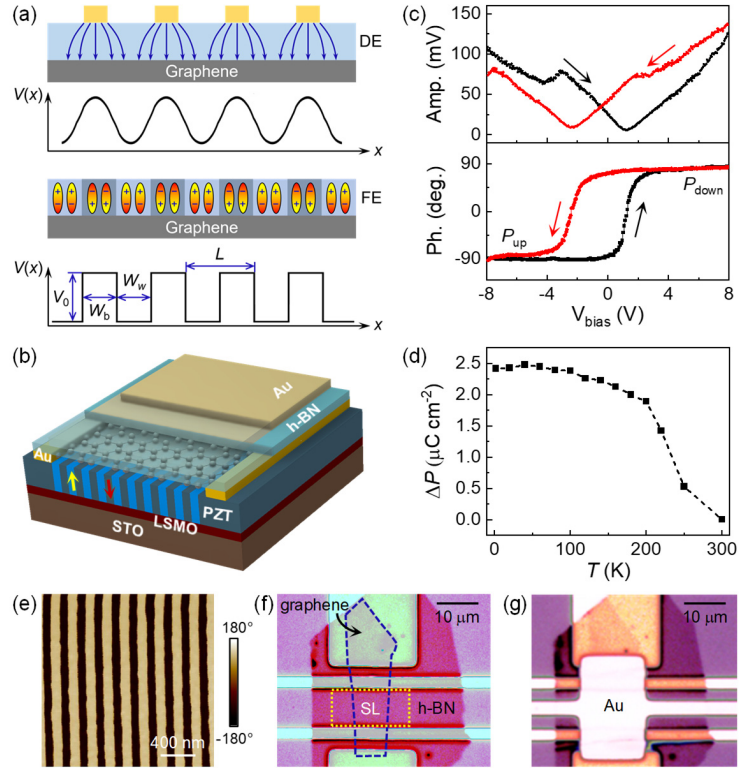

FIG. 1. (a) Schematics of 1D GSLs imposed via modulated dielectric (DE) gate (top) and ferroelectric (FE) domain structure (bottom), and the corresponding potential profiles $V(x)$. (b) Sample schematic. (c) PFM amplitude (top) and phase (bottom) switching hysteresis of 50 nm PZT/10 nm LSMO on STO. (d) Temperature-dependence of polarization for PZT. (e) PFM phase image of stripe domains with period of $L$ = 205 nm on PZT. (f) Optical image of a h-BN/graphene stack transferred on prepatterned domain structure of PZT. The dashed (dotted) lines highlight the graphene edges (SL region). (g) Optical image of the same sample after depositing Au top-gate electrode.

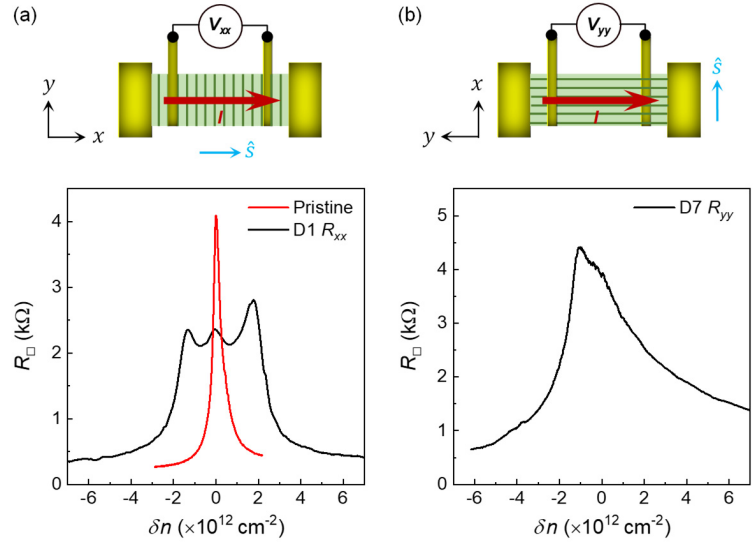

FIG. 2. (a) $R_\square(\delta n)$ for pristine graphene and sample D1. (b) $R_\square(\delta n)$ for sample D7. Upper panels: schematics of GSL configurations.

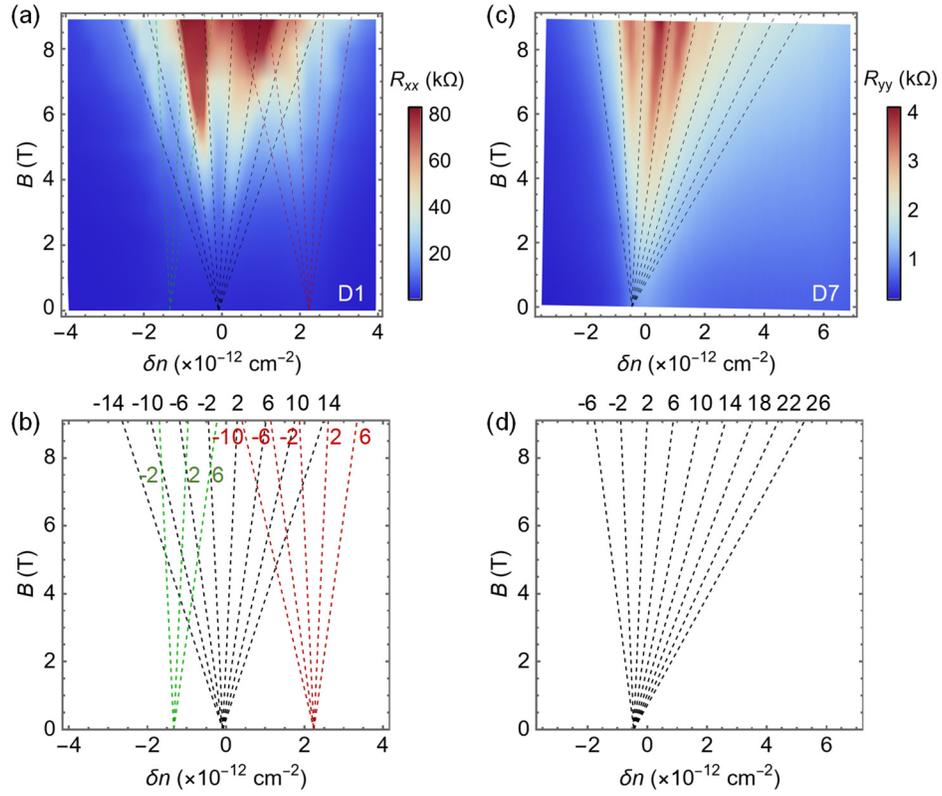

FIG. 3. (a) $R_{xx}$ vs. $\delta n$ and $B$ for sample D1. The dashed lines illustrate the modeled Landau fans emanating from the original and satellite DPs. (b) Modeled Landau fans in (a) with the filling factors labeled. (c) $R_{yy}$ vs. $\delta n$ and $B$ for sample D7. The dashed lines illustrate the modeled Landau fan emanating from the original DP. (d) Modeled Landau fan in (c) with the filling factors labeled.

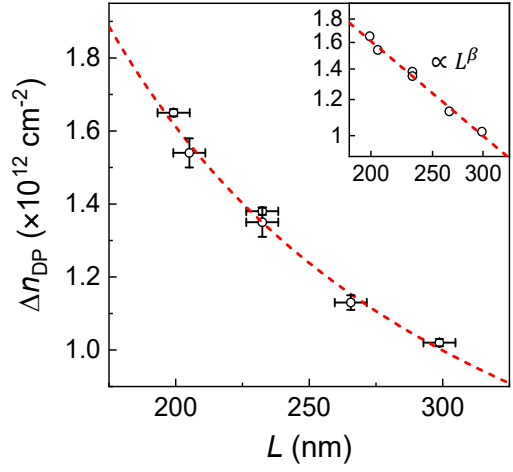

FIG. 4. $\Delta n_{\text{DP}}$ *vs.* $L$ for samples D1-D6 with a fit to Eq. 1 (dashed line) in linear and (inset) semi-log scales. The error bars for $\Delta n_{\text{DP}}$ and $L$ are deduced from the variation in h-BN thickness (SM Table S1) and DW roughness (SM Section 1), respectively [35].

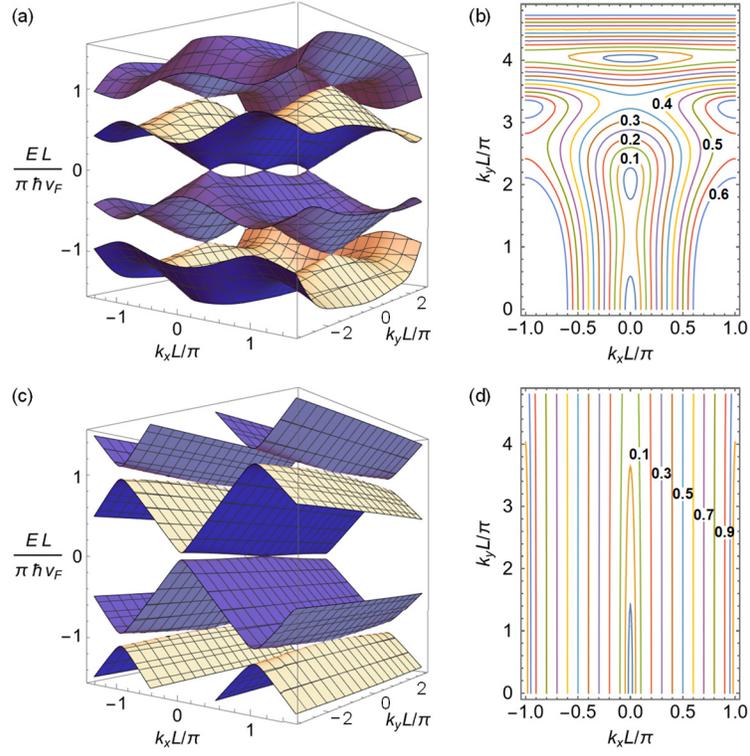

FIG. 5. (a) Calculated band structure at $u = 9\pi$ and $W_w = W_b$, and (b) its dimensionless energy contour plot of conduction band. (c) Calculated band structure at $u = 90\pi$ and $W_w = W_b$, and (d) its dimensionless energy contour plot of conduction band. The dimensionless energy values are labeled in (b) and (d).

**Transport Anisotropy in One-Dimensional Graphene Superlattice in the High Kronig-Penney Potential Limit (Supplemental Material)**

Tianlin Li,[1] Hanying Chen,[1] Kun Wang,[1] Yifei Hao,[1] Le Zhang,[1] Kenji Watanabe,[2] Takashi Taniguchi,[3] Xia Hong[1,*]

[1] *Department of Physics and Astronomy and Nebraska Center of Materials and Nanoscience, University of Nebraska, Lincoln, Nebraska 68588, USA*

[2] *Research Center for Electronic and Optical Materials, National Institute for Materials Science, 1-1 Namiki, Tsukuba 305-0044, Japan*

[3] *Research Center for Materials Nanoarchitectonics, National Institute for Materials Science, 1-1 Namiki, Tsukuba 305-0044, Japan*

[*] Email: xia.hong@unl.edu

1. Growth and Characterization of $PbZr_{0.2}Ti_{0.8}O_3$ Thin Films
2. Pyroelectric Effect in $PbZr_{0.2}Ti_{0.8}O_3$ Thin Films
3. Fabrication of 1D Graphene Superlattice
4. Robustness of Ferroelectric Domains Against Electrical Measurements
5. Emergent Extra Dirac Points and Effect of Temperature
6. Magnetotransport Results of 1D Graphene Superlattices
7. Analysis of Extra Dirac Point Index and $\Delta n_{DP}$

## 1. Growth and Characterization of PbZr$_{0.2}$Ti$_{0.8}$O$_3$ Thin Films

To design the one-dimensional (1D) graphene superlattice (GSL), we work with epitaxial 50 nm PbZr$_{0.2}$Ti$_{0.8}$O$_3$ (PZT) films deposited on 10 nm La$_{0.67}$Sr$_{0.33}$MnO$_3$ (LSMO) buffered (001) SrTiO$_3$ (STO) substrates. The PZT and LSMO layers are deposited *in situ* using off-axis radio-frequency magnetron sputtering, with the growth temperature of 500 °C and process gas of 120 mTorr (Ar:O$_2$ = 2:1) for PZT and the growth temperature of 650 °C and process gas of 150 mTorr (Ar:O$_2$ = 2:1) for LSMO. Figure S1(a) shows the x-ray $\theta$-$2\theta$ scan taken on the PZT/LSMO heterostructure, which reveals (001) growth for both PZT and LSMO layers with no impurity phases. The *c*-axis lattice constants for PZT and LSMO are 4.19 Å and 3.85 Å, respectively. We obtain the PZT thickness by fitting the Laue oscillations around the (001) Bragg peak of PZT [Fig. S1(b)].

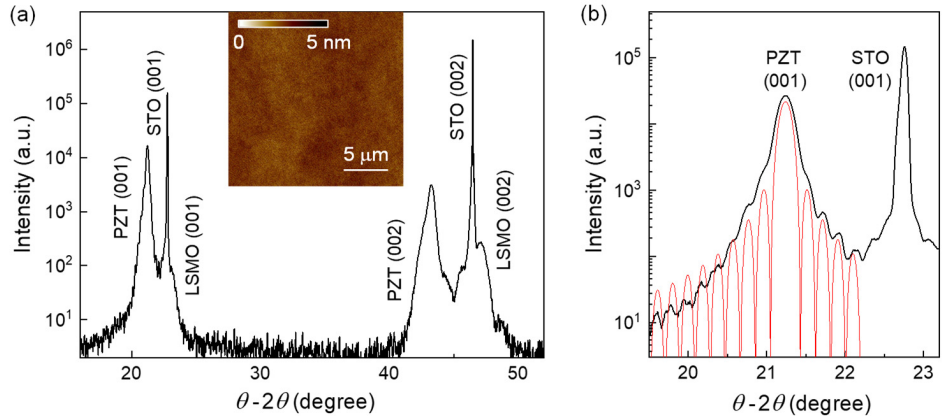

FIG. S1. (a) XRD $\theta$-$2\theta$ scan and (inset) AFM topography image taken on a 50 nm PZT/10 nm LSMO heterostructure deposited on (001) STO substrate. (b) XRD $\theta$-$2\theta$ scan around the (001) Bragg peak of PZT with a fit to the Laue oscillations.

We characterize the surface morphology and carry out conductive atomic force microscopy (AFM) and piezoresponse force microscopy (PFM) studies using a Bruker Multimode 8 AFM. The PZT films possess smooth surface morphology with a typical root-mean-square roughness of about 4 Å [Fig. S1(a) inset]. The conductive AFM and PFM measurements are performed using Pt/Ir-coated probes (Bruker SCM-PIT-V2). For domain writing, a DC bias ($V_{bias}$) larger than the coercive voltage of PZT is applied between the AFM tip and the bottom LSMO layer. Domain imaging is taken with an AC driving voltage of 0.4-0.8 V, which is lower than the coercive voltage, close to one of the resonant frequencies of the cantilever (315±20 kHz). The coercive voltages are determined from the PFM switching hysteresis [Fig. 1(c) in the main text].

Ferroelectric domain walls (DWs) can be viewed as elastic manifolds in disordered potential and are thus intrinsically roughened due to disorder pinning and thermal fluctuation. The DW roughness can introduce uncertainty to the superlattice period. To assess this effect, we create straight DWs in the PZT film [Fig. S2(a)] and convert the PFM phase image into a binary image, from which we identify the DW position [Fig. S2(b)]. We then calculate the root-mean-square of DW position with respect to the straight configuration and deduce a DW roughness of about 6 nm.

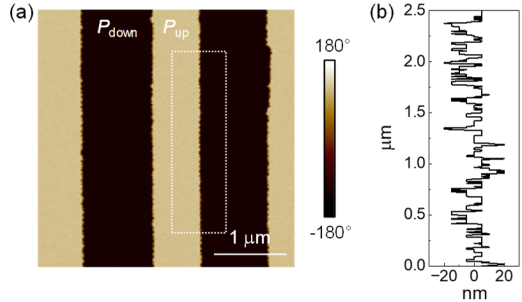

FIG. S2. (a) PFM phase image of stripe domains in PZT, with (b) the domain wall extracted from the boxed region in (a).

## 2. Pyroelectric Effect in PbZr$_{0.2}$Ti$_{0.8}$O$_3$ Thin Films

To determine the temperature dependence of polarization in PZT, we fabricate a graphene Hall bar device on PZT [Fig. S3 inset] and characterize the polarization doping induced carrier density in graphene. Figure S3 shows the Hall resistance $R_{xy}$ as a function of magnetic field $B$ at different temperatures. We extract the Hall coefficient $R_H = 1/ne$ from the slope of $R_{xy}(B)$ and calculate the carrier density $n$. From $n(T)$, we deduce the pyroelectric change of PZT polarization with respect to the room temperature value $\Delta P = |P - P\,(300\text{ K})| = e|n - n\,(300\text{ K})|$.

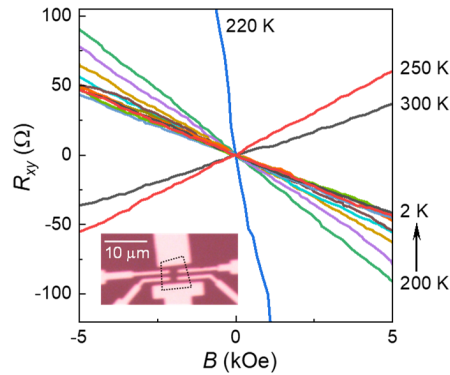

FIG. S3. $R_{xy}$ vs. $B$ at various temperatures taken on a graphene Hall bar device fabricated on PZT. Along the arrow: 200, 180, 160, 140, 120, 100, 80, 60, 40, 20, and 2 K. Inset: Optical image of the device. The dotted lines highlight the graphene boundary.

## 3. Fabrication of 1D Graphene Superlattice

To fabricate the 1D GSL, we define electrodes on the PZT films in the four-point geometry via laser writing followed by e-beam evaporation of Cr/Au (2 nm/10 nm). The entire region between the two voltage probes (about 17 μm by 11 μm) are first uniformly poled to the $P_{up}$ state by applying $V_{bias}$ = -6 V. Periodic stripes of $P_{down}$ domains are then written within the prepatterned $P_{up}$ background using the Bruker NanoMan function ($V_{bias}$ = +4.8, writing velocity of 12 μm/s). We fix the width of the $P_{down}$ stripes to about 100 nm. The SL period $L$ is 200 to 300 nm. Figure S4(a) shows the PFM image of the SL region of sample D1 with $L$ = 205 nm, which hosts 43 periodic stripe domains. The close-up PFM amplitude and phase images of the boxed region in Fig. S4(a) are shown in Fig. S4(b) and Fig. 1(e), respectively.

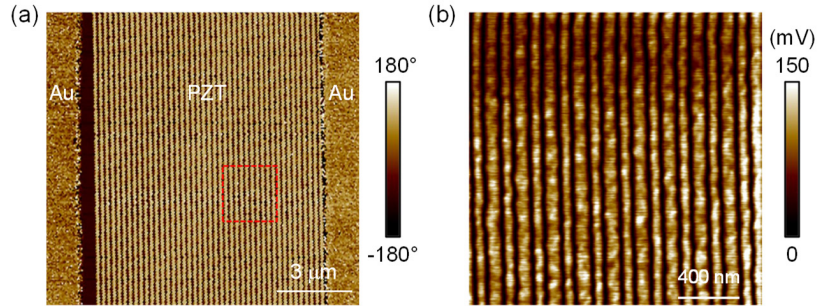

FIG. S4. (a) PFM phase image of periodic stripe domains written in PZT between a pair of Au electrodes. (b) PFM amplitude image of the boxed region in (a). The corresponding PFM phase image is shown in Fig. 1(e) in the main text.

After domain writing, the PZT substrates are ultrasonically cleaned in acetone, 2-propanol, and deionized water. Pre-prepared h-BN/monolayer graphene stacks are transferred onto PZT using the dry transfer approach, with the graphene channels in direct contact with the periodic domain structures and the Au electrodes. The interfacial air bubbles are then removed by vacuum annealing at about 100 °C for 2 hours.

## 4. Robustness of Ferroelectric Domains Against Electrical Measurements

Figure S5(a) shows the PFM phase image of stripe domains on PZT, upon which we fabricate the graphene field effect sample [Fig. S5(b)]. After electrical measurements at 2 K, we peel off the Au/h-BN/graphene stack [Fig. S5(c)]. As shown in Fig. S5(d), the stripe domains remain unchanged, confirming that the ferroelectric domain induced SL modulation is robust against the sample fabrication and characterization processes.

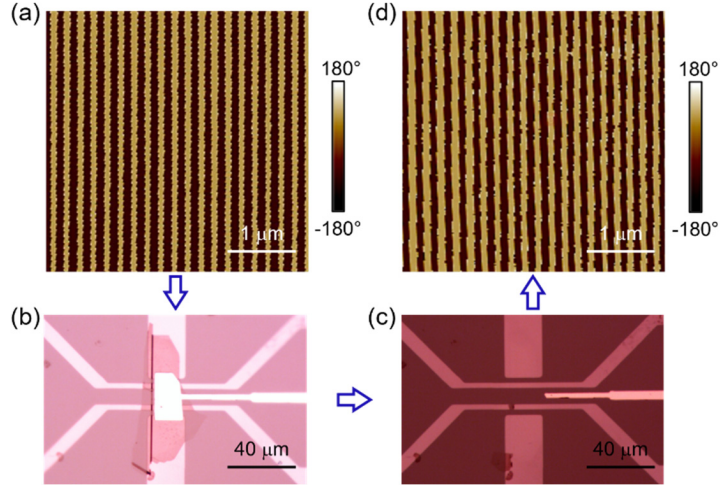

FIG. S5. (a) PFM phase image of stripe domains on PZT with period $L = 200$ nm. (b) Optical image of a 1D GSL sample fabricated on the domain structure shown in (a). (c) Optical image of the same sample area after the Au/h-BN/graphene stack is peeled off, and (d) the corresponding PFM phase image of the stripe domain region.

## 5. Emergent Extra Dirac Points and Effect of Temperature

Figure S6 shows the $R_\square$ as a function of h-BN top-gate voltage $V_g$ at 2 K for a pristine graphene sample [Fig. S6(a)] and two GSL samples with current parallel with [Fig. S5(b)] and perpendicular to the SL vector $\hat{s}$ [Fig. S6(c)].

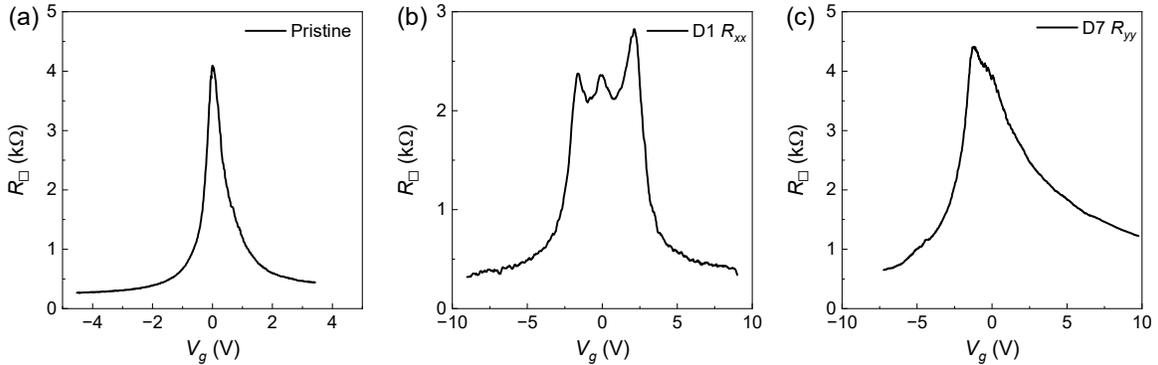

FIG. S6. $R_\square(V_g)$ for (a) pristine graphene with h-BN thickness $d = 32.7$ nm, (b) sample D1 with $d = 25.3$ nm and SL period $L = 205$ nm, and (c) sample D7 with $d = 24.5$ nm and SL period $L = 199$ nm.

Figure S7 shows $R_{xx}(V_g)$ of sample D1 measured at different temperatures (1.8-10 K). The positions of the extra Dirac points (DPs) do not vary significantly, consistent with the less than 5% polarization change of PZT below 100 K induced by the pyroelectric effect [Fig. 1(d)]. Above 10 K, the leak current through h-BN increases significantly, which reduces the workable $V_g$-range.

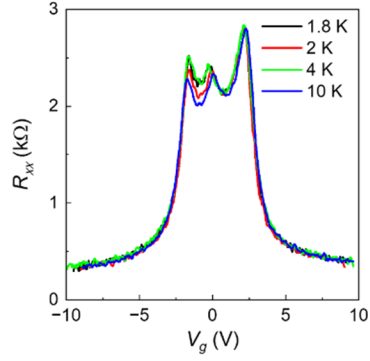

FIG. S7. $R_{xx}$ vs. $V_g$ of sample D1 at 1.8 K, 2 K, 4 K, and 10 K.

## 6. Magnetotransport Results of 1D Graphene Superlattices

Figure S8(a)-(b) shows $R_{xx}(V_g)$ of sample D1 measured at different magnetic fields. In magnetic field, $R_{xx}$ exhibits complex oscillatory features due to the superposition of Shubnikov de Haas oscillations originating from three DP branches (DP indices $N$ = -1, 0, 1). The data can be well captured by the Landau fans emanating from these DPs, as shown in Fig. 3(a) in the main text. Such multiple-Landau-fan feature has also been observed in other 1D GSL samples [Fig. S8(c)-S8(d)].

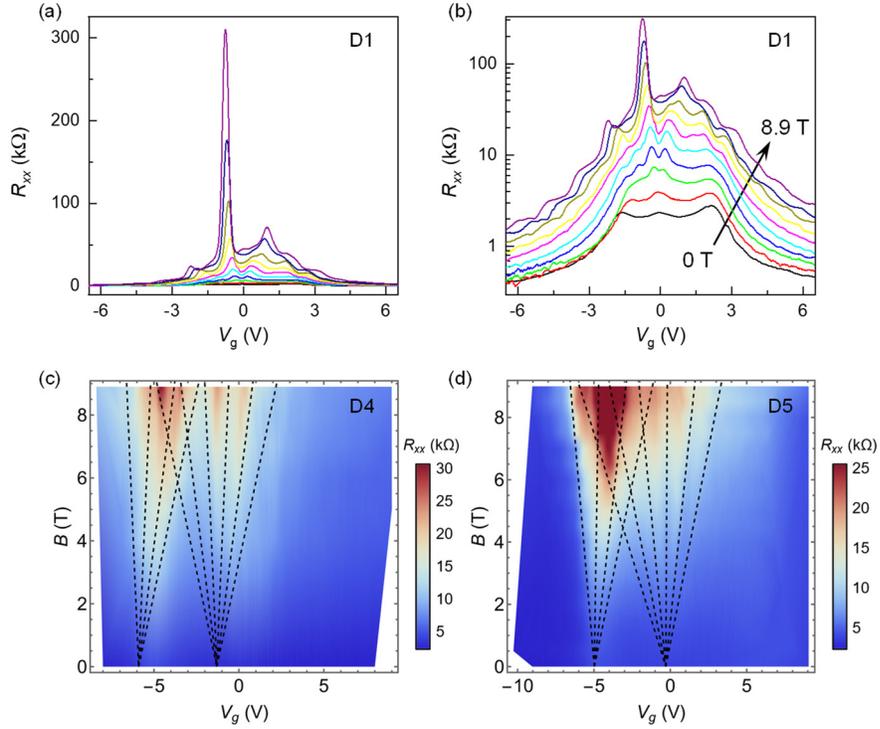

FIG. S8. (a)-(b) $R_{xx}(V_g)$ of sample D1 at 2K in linear (a) and semi-log (b) plots at $B$ = 0, 1, 2, 3, 4, 5, 6, 7, 8, and 8.9 T. (c)-(d) $R_{xx}$ vs. $V_g$ and $B$ for samples D4 (c) and D5 (d). The dashed lines illustrate the modeled Landau fans emanating from the original and hole branch satellite DPs.

## 7. Analysis of Extra Dirac Point Index and $\Delta n_{DP}$

We report the $R_{xx}$ results of six 1D GSL samples (samples D1-D6) with the SL period $L$ varying from 200 to 300 nm. Figure S9 shows the $R_{xx}$ vs. top-gate induced electron doping $\delta n$ for D1-D6, where we adopt local peak fits to determine the position and indices of the extra DPs. Figure S10 shows the global multiple-peak fits combining contributions from all peaks, which successfully captures the main features of the experimental results (Fig. S9).

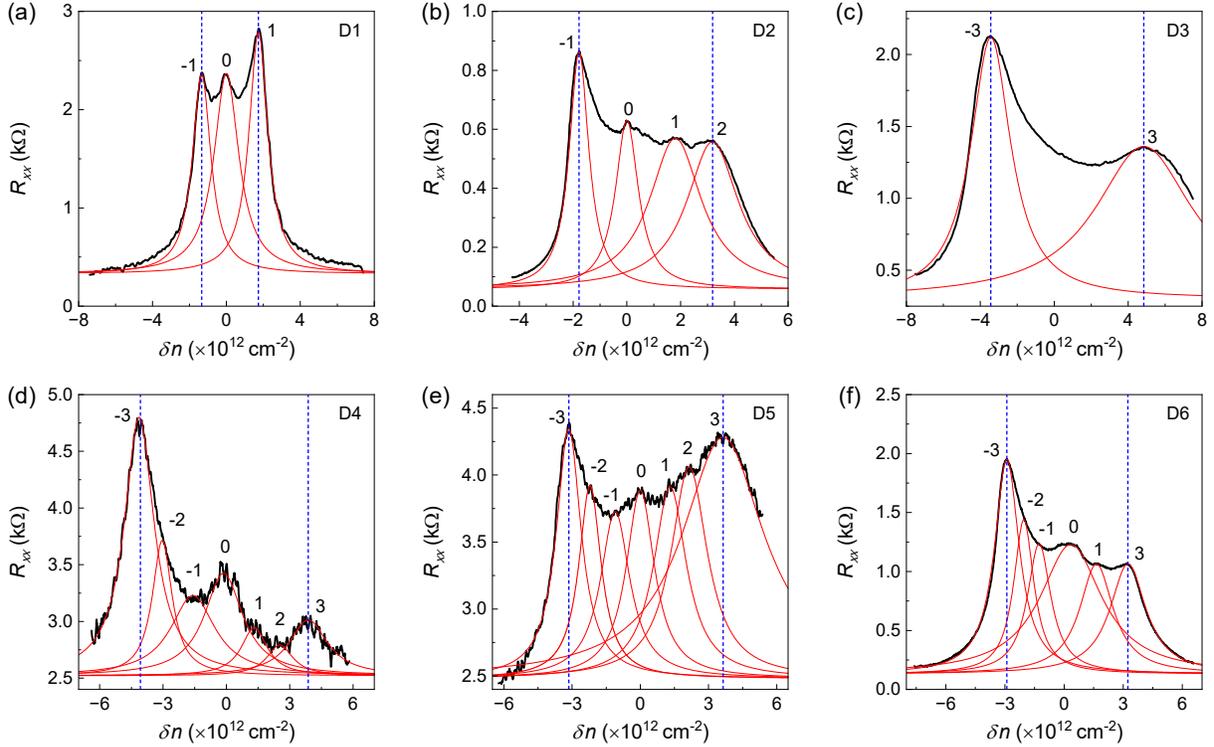

FIG. S9. $R_{xx}(\delta n)$ for samples D1-D6 at 2 K. The red lines are Lorentz fits to local peaks for each extra DP, with the index $N$ labeled. The blue dashed lines indicate the peaks used for calculating $\Delta n_{DP}$.

As discussed in the main text, the large Kronig-Penney potential imposed by ferroelectric domains leads to enhanced density of states at the band crossing point, which broadens the resistance peaks. At low doping level, due to the suppressed mean free path and hence insufficient band reconstruction, peaks of some low order ($N$) extra DPs cannot be clearly resolved. The order $N$ of the resolved DPs are determined by the following assumptions:
1. The $N = 0$ DP is close to $n = 0$.
2. The electron and hole branches should have symmetric distributions of the DPs.

3. With small variation in the SL vector, the distribution for the DP doping level should not change abruptly.

In Fig. S9-S10, we label the identified indices $N$ for the DPs. Two most prominent peaks at high doping levels (labeled by the blue dashed lines) are selected to calculate the average carrier density interval between two consecutive DP ($\Delta n_{DP}$). Please note assumption 1 does not affect the deduced value of $\Delta n_{DP}$.

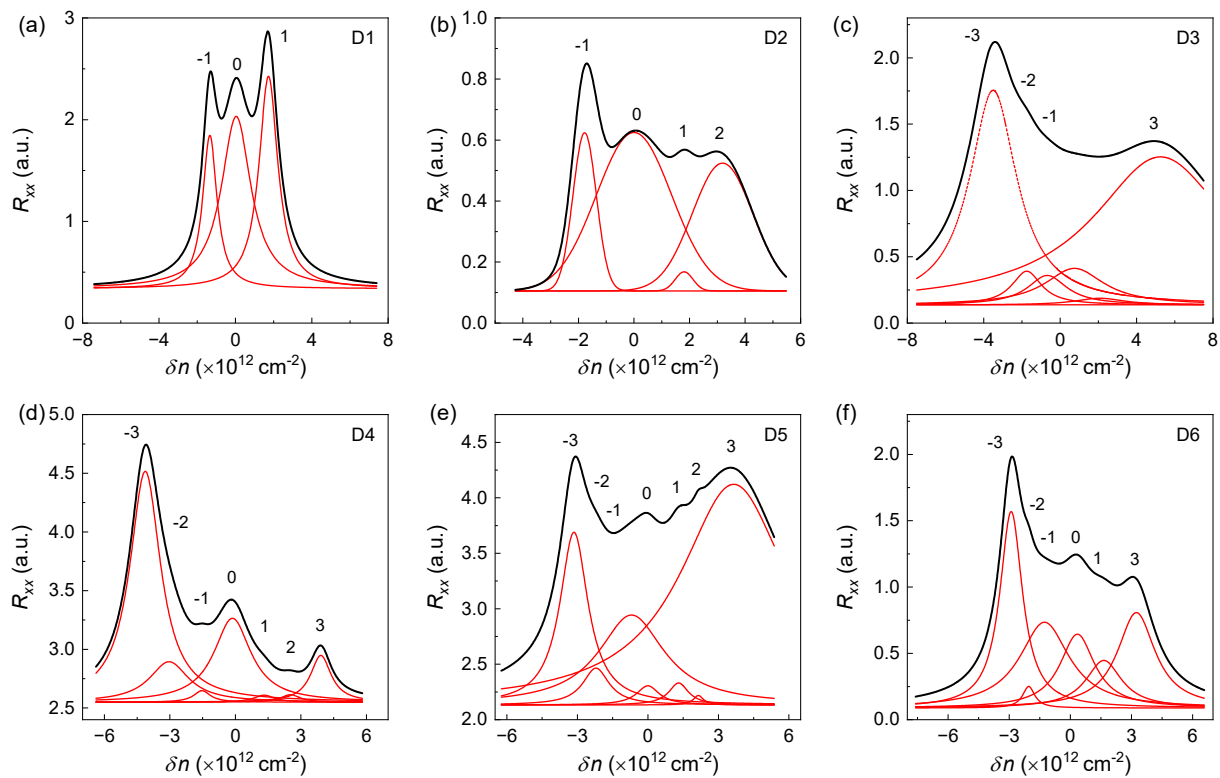

FIG. S10. Global fits of $R_{xx}(\delta n)$ for D1-D6 at 2 K. The red curves are fits of the individual peaks. The black curves are the sum of the red curves.

Table S1 summarizes the sample parameters of six 1D GSL samples with the $R_{xx}$-configuration (D1-D6) and one 1D GSL sample with $R_{yy}$-configuration (D7).

| Sample # | $L$ (nm) | $d$ (nm) | $V_{DP}$ (V) | $N$ | $\Delta n_{DP}$ ($\times 10^{12}$ cm$^{-2}$) |
|---|---|---|---|---|---|
| $R_{xx}$-configuration | | | | | |
| D1 | 205±6 | 25.3±0.7 | -1.63 | -1 | 1.54±0.04 |
| | | | 2.11 | 1 | |
| D2 | 200±6 | 40.1±0.3 | -3.43 | -1 | 1.65±0.01 |
| | | | 6.15 | 2 | |
| D3 | 233±6 | 16.6±0.5 | -2.74 | -3 | 1.38±0.04 |
| | | | 3.87 | 3 | |
| D4 | 233±6 | 29.2±0.4 | -5.94 | -3 | 1.35±0.02 |
| | | | 5.45 | 3 | |
| D5 | 267±6 | 32.9±0.4 | -4.98 | -3 | 1.13±0.01 |
| | | | 5.76 | 3 | |
| D6 | 300±6 | 30.0±0.4 | -4.19 | -3 | 1.02±0.01 |
| | | | 4.68 | 3 | |
| $R_{yy}$-configuration | | | | | |
| D7 | 199±6 | 24.5±0.3 | N/A | N/A | N/A |

Table S1. A summary of parameters for 1D GSL samples, including superlattice period $L$, top-layer h-BN thickness $d$, $V_g$ position of the selected extra DP $V_{DP}$, the corresponding index $N$ of Brillouin zone folding, and the deduced average carrier density interval between consecutive DPs $\Delta n_{DP}$. For $V_{DP}$ and $N$, "-" corresponds to the hole branch and "+" corresponds to the electron branch.